\documentstyle{article}
\title{ \bf Square Gravity
}
\author{ {\it C.F. Baillie} \\
         Dept. of Computer Science\\
         University of Colorado\\
         Boulder, CO 80309, USA\\
         \\
         and \\
         \\
         {\it D.A. Johnston} \\
         Dept. of Mathematics\\
         Heriot-Watt University\\
         Riccarton\\
         Edinburgh, EH14 4AS\\
         Scotland\\
         }
\date { 18 June 1995 }         %if no \date --> current date
\begin{document}
  \maketitle
%-----------------------------------------------------------------------
                      {\Large
                      \begin{abstract}
%-----------------------------------------------------------------------
%
We simulate the Ising model on dynamical quadrangulations
using a generalization of the flip move 
for triangulations with two
aims: firstly, as a confirmation of universality
for the KPZ/DDK exponents of the Ising phase transition,
worthwhile in view of some recent surprises with
other sorts of dynamical lattices; secondly,
to investigate the transition of the Ising {\it anti}-ferromagnet
on a dynamical loosely packed (bipartite) lattice. In the latter case
we show that it is still
possible to define a staggered magnetization and observe the
anti-ferromagnetic analogue of the transition.
\\
\\
Submitted to Physics Letters B.
%
%-----------------------------------------------------------------------
                        \end{abstract} }
%-----------------------------------------------------------------------
%
  \thispagestyle{empty}
%
%***********************************************************************
%
  \newpage
%
%-----------------------------------------------------------------------
                  \pagenumbering{arabic}
%-----------------------------------------------------------------------
\section{Theoretical Background}

The behaviour of various spin models on dynamical triangulations
and their dual planar $\phi^3$ graphs has been extensively investigated
both analytically \cite{1,1a} and numerically \cite{2,2a}
recently because of their interest for 
string theory and $2d$ gravity.
The sum over triangulations
or graphs in such models serves as a 
discretization of the sum over metrics
in the continuum theories. In simulations the space
of triangulations or their duals is generally sampled by performing
local ``flip'' moves \cite{2b}, 
though more recently 
cluster-like moves for the geometry (``baby-universe
surgery'') have been proposed \cite{2c}. 
Considering
multiple copies of spin models, or alternatively large $Q$ Potts models,
has led to considerable progress in understanding the 
$c>1$ regime in $2d$ quantum gravity \cite{3,3a}.

The microcanonical
partition function for an Ising model
on some ensemble of graphs with $n$ vertices is given by
\begin{equation}
Z_n(\beta, H) = \sum_{G^n} \sum_{\{ \sigma \}} \exp \beta \left( \sum_{<ij>} G^n_{ij} \sigma_i \sigma_j
+ H \sum_i \sigma_i \right)
\label{e01}
\end{equation}
where $G^n_{ij}$ is the connectivity matrix for the graph 
$G^n$.
For the particular case 
of the Ising model on planar $\phi^3$ and $\phi^4$ graphs
the solution
proceeds by noting the equivalence of the grand canonical
partition function arising from equ.(\ref{e01}) and the free
energy of an $N \times N$ two matrix model \cite{1}.
In the $\phi^4$ case in zero external field this is
\begin{equation}
F_{\phi^4} (c, g, H) = {1 \over N^2} \log \left( \int d^{N^2} M_1 d^{N^2} M_2 \exp  - tr {N \over g} \left( {M_1^2 \over 2} + {M_2^2 \over 2} - c M_1 M_2
+ { 1 \over 4} M_1^4 + {1 \over 4} M_2^4 \right) \right)
\label{e02}
\end{equation} 
and for $\phi^3$ graphs 
\begin{equation}
F_{\phi^3} (c, g, H) = {1 \over N^2} \log \left( \int d^{N^2} M_1 d^{N^2} M_2 \exp  - tr {N \over g} \left( {M_1^2 \over 2} + {M_2^2 \over 2} - c M_1 M_2
+ { 1 \over 3} M_1^3 + {1 \over 3} M_2^3 \right) \right)
\label{e03}
\end{equation} 
where $c = \exp(- 2 \beta)$ introduces the temperature of the spins
into the theory. In the planar limit $N \rightarrow \infty$ it is possible
to solve the model exactly using orthogonal polynomial or saddle point
methods and derive the full set of critical exponents, which are
in agreement with those calculated from continuum conformal field theory
\cite{1a}.

It is expected that the Ising model on any dynamical
polygonization or its dual will display the KPZ/DDK critical behaviour
so long as further couplings are not tuned to give a multicritical 
point. This has been demonstrated both analytically and numerically
for various variants of $\phi^3$ graphs where self-energy
diagrams and tadpoles are either excluded or not. Although the critical
temperature changes, the critical exponents are unaffected.
There have, however, been some recent surprises
with other discretizations:
with $2d$ Regge calculus the Onsager exponents were observed
\cite{04}; and with apparently ``flat'' models 
the KPZ/DDK exponents were measured \cite{05}.
This increases the incentive for performing
an explicit check on the behaviour of various possible
discretizations.

It is possible to write down explicitly a matrix model
for the Ising model on a dynamical triangulation (ie with the spins
residing at the vertices of the triangle, rather than in the centre
as is effectively the case for $\phi^3$ graphs). The
action of this model is given by
\cite{4}
\begin{equation}
U_{DTRS} =  { N \over g} tr \left( {1 \over 2 \cosh (\beta) (1 + c^*)} S^2 + { 1 \over 2 \cosh ( \beta) ( 1 - c^*)} D^2
+ S^3 / 3 + S D^2 \right)
\label{e04}
\end{equation}
where $c^* = (1 - c) / ( 1 + c)$ is the transformation
induced in $c$ by the standard duality transformation on $\beta$
and $S$ and $D$ are still $N \times N$ hermitian matrices
representing edges with the
same and different spins at the end respectively.
This model may be transformed into a form
equivalent to the
``$O(1)$'' representation of the Ising model
on $\phi^3$ graphs
which is discussed in \cite{5}
\begin{equation}
U_{\phi^3} = { N \over g} tr \left( D^2 S + { S^3 \over 12} - {c^* \over 2} S^2 + { (3 c^* - 1) (1 + c^*)\over 4} S \right)
\label{e05}
\end{equation}
and shown to have Ising critical behaviour at the dual of the 
$\phi^3$ model transition temperature. 
Working directly on the triangulations with $U_{DTRS}$
thus provides an explicit confirmation of duality and 
a further check on the universality
of the exponents.

We can
carry out a strong coupling
expansion whatever the lattice our Ising model inhabits 
\begin{equation}
Z_{V}(\beta, 0) = \sum_{G^V}  2^{V} (\cosh ( \beta) )^{E} \sum_{loops} (\tanh (\beta))^{length}
\label{e06}
\end{equation} 
where the loop sum is restricted to edges that are only traversed
once. This makes it clear that we should only expect to see
an {\it anti}-ferromagnetic transition in the Ising
model on lattices where only even loops are possible,
because only here do we have a $\beta \rightarrow -\beta$ symmetry.
Thus, for instance, the honeycomb lattice in $2d$ displays an
antiferromagnetic transition whereas the triangular lattice does not.
It is not entirely trivial to conclude that such transitions
would survive a coupling to $2d$ gravity, which in effect
introduces annealed connectivity disorder into the lattices.
Both the analogue of the honeycomb lattice -- dynamical $\phi^3$ graphs,
and the triangular lattice -- dynamical triangulations, fail
to display an antiferromagnetic transition because both odd and even
loops are present.

Graphs with only even loops can be constructed in
general by considering complex matrices \cite{6}, but we take a slightly
different tack here and consider the Ising model
on a dynamical quadrangulation, where the action is
given by
\begin{equation}
U_{DQRS} = {N \over g} tr \left( {1 \over 2} S^2  + { 1 \over 2} D^2 + {g_1 \over 4} S^4 + {g_2 \over 4} D^4 + {g_3 \over 2} ( S D S D + 2 S^2 D^2 ) \right)
\label{e16}
\end{equation}
with
\begin{eqnarray}
g_1 &=& \cosh^2 ( \beta) ( 1 + c^*)^2 \nonumber \\
g_2 &=& \cosh^2 ( \beta) ( 1 - c^*)^2 \nonumber \\
g_3 &=& \cosh^2 ( \beta) ( 1 + c^*) (1 - c^*).
\label{e17}
\end{eqnarray}
This model can be shown to have a transition at $c=3/5$
by transforming the action back to a form resembling that
for the Ising model on $\phi^4$ graphs.
With some scalings we can rewrite equ.(\ref{e17}) as
\begin{equation}
U_{DQRS} = {N \over g} tr \left( {1 \over 2} S^2  + { 1 \over 2} D^2 + {1 \over 4 ( 1 - c^*)^2} S^4 + {1\over 4 ( 1 + c^*)^2} D^4
 + {1 \over 2 ( 1 - c^*) ( 1 + c^*)} ( S D S D + 2 S^2 D^2 ) \right).
\label{e19}
\end{equation}
The $\phi^4$ model of equ.(\ref{e02}) 
is recast by
setting
$A = ( M_1 + M_2 ) / \sqrt{2}$, $B = (M_1 - M_2) / \sqrt{2}$ and then rescaling $A \rightarrow \sqrt{2} A / \sqrt{ 1 - c}$, $B \rightarrow \sqrt{2} B / \sqrt{ 1 + c}$,
$g \rightarrow 2 g $. 
This gives 
\begin{equation}
U_{\phi^4} = {N \over g} tr \left( {1 \over 2} A^2  + { 1 \over 2} B^2 + {1 \over 4 ( 1 - c)^2} A^4 + {1\over 4 ( 1 + c)^2} B^4
 + {1 \over 2 ( 1 - c) ( 1 + c)} ( A B A B + 2 A^2 B^2 ) \right).
\label{e20}
\end{equation}
which is identical in form to equ.(\ref{e19}). Given that $U_{\phi^4}$
produces an Ising transition at $c=1/4$, comparison with
equ.(\ref{e19}) shows that $U_{DQRS}$ produces an Ising transition
at $c^*=1/4$ (ie $c=3/5$), the quoted result.
Furthermore, the possibility of an antiferromagnetic transition is
manifested in equ.(\ref{e19}) too as $\beta \rightarrow - \beta$
(ie $c^* \rightarrow - c^*$) is obviously a symmetry of the action
upon exchange of $S$ and $D$. This is {\it not} the case for
the original $\phi^4$ action, even in its modified form 
in equ.(\ref{e20}),
as $c \rightarrow 1/c$ when $\beta \rightarrow - \beta$. 

The matrix model considerations would thus suggest that
the Ising model on dynamical quadrangulations displays
an antiferromagnetic transition with the same critical exponents
as the ferromagnetic case. This is not
a totally vacuous observation, as the antiferromagnetic
transition essentially operates on two decoupled sets 
of Ising spins, so it is not entirely obvious that one should
see the exponents for a one Ising model transition coupled to gravity
rather than those for two Ising models. 
To check our observations with a simulation requires the definition
of a flip move in the space of quadrangulations, that possesses
all the nice ergodicity properties of the standard triangle
flip moves. In \cite{4} we baldly presented one candidate move,
shown again in Fig.1, without any discussion of its
derivation or ergodicity properties. We now remedy this lacuna.

The easiest approach to take is to break our quadrangulation
into a triangulation and reconstruct the possible moves from that.
We shall consider undecorated (no spin) quadrangulations
first.
Not every triangulation can be paired off to give a
quadrangulation, we need one that admits a perfect matching.
A matrix model to do just this 
(in the dual $\phi^4/\phi^3$ picture) was discussed in \cite{7}
\begin{equation}
Z = \int D^{N^2} \phi \; \exp - \left( {1 \over 2} \phi^2 - { 1 \over 4 N}
e^{\ln 2 - \mu} \phi^4 \right) = 
\int D^{N^2} \phi \;  D^{N^2}u \; \exp -\left( {1 \over 2}\phi^2
+ {1 \over 2} u^2 - { 1 \over \sqrt{N} } e^{ - \mu / 2} \phi^2 u \right)
\label{e21}
\end{equation} 
where both $\phi$ and $u$ are $N \times N$ hermitian matrices.
Integrating out $u$ on the right hand side gives back the single
matrix model on the left hand side. If we go back to the direct
lattice picture of triangles and squares the effect of the 
integration
in the above dual lattices is to pair off all triangles
into squares. If we look at the dual $\phi^4$ diagram to two adjoining 
squares it can be broken down, or perhaps
more correctly exploded, in
four possible ways with the insertion of $u$ propagators
as shown in Figs.2. The standard $\phi^3$ flip moves can be performed
subject to the constraint that every vertex must retain one $u$
and two $\phi$ edges, before collapsing the $u$ propagators
to obtain the final $\phi^4$ diagram. Performing
the allowed flips on Figs.2c,d leave the final $\phi^4$ diagram unchanged
and the results of the allowed flips on Figs2a,b are shown in Fig.3a
and Fig.3b. Just as in the $\phi^3$ flip move
both labellings of the central segment appear with equal
probability in both the diagrams, and just as for the $\phi^3$ flip
the distinction disappears when one returns to the direct lattice.

When decorating the lattice with spins it is best to think
of the spin at the exploded vertices living on the $u$
propagator and getting carried with it in any flip moves.
This avoids any potential conflict that might arise
when collapsing the diagram after performing the flips
if we actually created new spins on the split vertices.
The net result of all of this is that the flip moves deduced
by the procedure of exploding the
$\phi^4$ diagram into $\phi^3$ diagrams and carrying out
$\phi^3$ flips before reconstructing the $\phi^4$ diagram
are precisely those given in Fig.1. Both the moves appear with
equal probability. As we can trace the moves back to the 
$\phi^3$ picture (albeit with two sorts of propagators) the ergodicity
is assured.

A final subtlety to consider
before launching into the simulations proper is the
choice of class of quadrangulation. We restrict ourselves
for reasons of computational simplicity to non-degenerate
quadrangulations in which we do not allow such nasties as squares
touching on more than one side. In the dual $\phi^4$ picture this
corresponds to excluding bubbles on propagators, the ``setting-sun''
diagram and the one loop contribution to the $\phi^4$ vertex.
Recent simulations on $\phi^3$ graphs have shown, 
rather counterintuitively, that including the $\phi^3$ counterparts
of such diagrams actually improves the scaling behaviour and we hope
to return to this issue for quadrangulations in further work.

\section{Simulations}

We simulate graphs of size $N=100,225,400,900$ and $2500$ vertices.
In all cases the starting configuration was taken to be
a regularly quadrangulated torus for simplicity
(e.g. $2500 = 50 \times 50$). The topology of
the manifold on which the spin model lives is not expected to
change the KPZ/DDK exponents for the spin model transition, although
$\gamma_{string}$, which we do not measure here, will of course
depend on the topology. We conducted 10,000 equilibration sweeps
and 50,000 measurement sweeps for most of the lattice sizes
at each $\beta$ value simulated, with more sweeps on the larger
graphs and closer to the phase transition. 
For the ferromagnetic model we updated the spins
with the Wolff cluster algorithm, ensuring that on average
every spin was updated per sweep. The number of flip updates per sweep
was chosen to be equal to $N$, a rule of thumb that works well
for $\phi^3$ graphs and triangulations, and seems to do so here too.
Varying the number of flip updates by a factor of two in either direction
had little effect on the measured quantities. Although it is
possible to construct a cluster algorithm for the antiferromagnetic model
as well, we contented ourselves with an old-fashioned Metropolis update
in this case, again for simplicity as very high accuracy  was not
a primary requirement.

We measured all the standard thermodynamic quantities for the model:
the energy $E$, specific heat $C$, magnetization $M$, susceptibility $\chi$
and various cumulants. A short word is in order on the magnetization
and energy measurements for the antiferromagnetic model. The normalization
of the energy (ie the constant term in the hamiltonian) is chosen so that
the energy runs between $2$ and $4$ for both the ferromagnet and
antiferromagnet. In Fig.4 we plot the energy in both models. It is clearly
essentially identical, especially when the less efficient algorithm employed
for the antiferromagnet is taken into account. This provides the first confirmation 
of the matrix model result that the antiferromagnetic transition should be identical
to the ferromagnetic transition on dynamical quadrangulations.
We have to work slightly harder to demonstrate the equality of the 
magnetizations as the standard magnetization will clearly stay zero
in any antiferromagnetic transition. It is, however, still possible
to easily
define a staggered magnetization on the dynamical quadrangulation
because of our choice of starting configuration. We can assign an initial
sign or parity to each vertex in a ``checkerboard'' pattern consistently
on the initial regularly quadrangulated torus. The flip move
of Fig.1 maintains the parity of flipped links as is clear from
looking at the solid and open dots in the diagram, so we can define
a staggered magnetization as
\begin{equation}
M = { 1 \over N} \sum_{ij} (-1)^{i+j} \sigma_{ij}
\end{equation}
where $i,j$ labels the original 
horizontal and vertical coordinates of the spin in the starting torus.
Although our choice of a regular torus as
starting point makes the definition of the staggered magnetization
particularly easy, it should be stressed that {\it any} loosely
packed lattice could be checkerboarded in such a manner and that
the corresponding flip moves would not disrupt these assignments
or the definition of the staggered magnetization.
In Fig.5 we plot the standard magnetization for the
ferromagnet and the staggered magnetization for the antiferromagnet,
which are again very nearly equal until the inefficiency of the 
metropolis algorithm begins to bite in the low temperature phase.
Note the large error bars in this case; all the other error bars are 
smaller than the symbols.

Having convinced ourselves of the qualitative similarity of the
ferromagnetic and antiferromagnetic transitions from these figures
we now proceed to a more quantitative scaling analysis. We commence
by looking at the Binder's cumulant for the magnetization
\begin{equation}
U_M = 1  - { <M^4> \over 3 <M^2>}
\end{equation}
whose crossing point for different lattice sizes determines $\beta_c$.
The data for both the ferromagnet and antiferromagnet
give quite clean crossings that are consistent for the various
different size pairings, in contrast to our $\phi^3$ simulations
in \cite{2a} where we had to resort to a 
slightly more devious approach
to extract a reliable estimate for $\beta_c$.
We find $\beta_c = 0.398(2)$ for the ferromagnet and
$\beta_c = -0.40(2)$ for the antiferromagnet, where the larger error
in the latter is due to the poorer statistics of the metropolis
algorithm employed. Note that we have not found $\beta_c = -1/2 \log (3/5)
\simeq 0.255 \ldots$ as given by the matrix model because the matrix
model calculation implicitly assumes the inclusion of the degenerate
squares that we have expunged. It would be an interesting exercise
to calculate the expected $\beta_c$ for our variant analytically
to compare with the measurements above.

We can extract the combination $\nu D$ from $U_M$ in various ways.
One possibility is to look at the maximum slope of
$U_M$, which is expected to scale as
\begin{equation}
\max \left( { d U_M \over d \beta} \right) \ \simeq \ N^{1 / \nu D}.
\end{equation}
For the
ferromagnet we find that $1 / \nu D = 0.31(5)$ or $\nu D =3.2(5)$, where the
rather large error bar is accounted for by the effect of the smaller lattices.
Fitting to 
only the larger lattice sizes gives both a slightly smaller estimate
and a smaller error. If we accept the likelyhood of finding KPZ/DDK
rather than
Onsager exponents the hyperscaling relation  $\alpha = 2 - \nu D$ would
give us $\nu D = 3$ rather than $\nu D = 2$.
The above value is clearly in better agreement with the KPZ/DDK result
than the Onsager result.
An alternative is to look at the scaling of the slope at
the crossover point, which gives slightly a slightly better fit of $
1 / \nu D =0.31(3)$ or $\nu D =3.2(3)$, consistent with the above values.
The results for the antiferromagnet are very similar, again with larger errors
due to the poorer statistics. Given $\nu D$ and $\beta_c$
we can now proceed to a finite size scaling analysis of some of the other
exponents. In the discussion which follows, the results are
for the ferromagnet unless explicitly indicated otherwise.

Although a fit to the finite size scaling relation for the specific
heat is not particularly illuminating because of the extra adjustable
constant $A$ that is present
\begin{equation}
C \simeq A + B N^{\alpha / \nu D}
\end{equation}
it is consistent with the values of $\nu D$ coming from the cumulant.
The specific heat for the ferromagnet plotted in Fig.6 (that for
the antiferromagnet is identical) clearly shows the expected cusp, the 
weak growth of the peak for small lattice sizes dying away as the size
is increased. Working in lexicographic order, we now turn our attention to
the exponent $\beta$
\begin{equation}
M \simeq M_0 N^{- \beta / \nu D}
\end{equation}
which appears to give rather poor fits on modestly sized $\phi^3$
lattices \cite{2a}, although the use of a triangulation \cite{2}
or the inclusion of tadpoles and self-energies greatly improves matters
\cite{3}. With quadrangulations we find that we have maintained
our traditionally atrocious fits to $\beta$, finding 
$\beta / \nu D = 0.075(2)$, which translates
to $\beta=0.24(3)$ rather than the KPZ/DDK value of 0.5.
The fits to $\gamma$ from the susceptibility $\chi$
\begin{equation}
\chi \simeq \chi_0 N^{\gamma / \nu D}
\end{equation}
are much more satisfactory, however. We find $\gamma / \nu D = 0.72(2)$,
which using our best fitted value of $\nu D = 3.2(3)$ gives the estimate
of $\gamma=2.3(3)$ in quite good agreement with the KPZ/DDK value of 2.
We can see clearly the sharp divergence in $\chi$ in Fig.7.
A similar story holds for the antiferromagnetic simulations:
a good fit
to the KPZ/DDK value for 
$\alpha / \nu D$ is still possible because of the 
constant term, we find $\beta / \nu D = 0.078(2)$ -- again too small, 
and $\gamma / \nu D = 0.73(7)$.

The lattices we simulate are rather modestly sized,
but it is still possible to do a direct fit to the dependence
of $C,M$ and $\chi$ on the 
reduced temperature $t = |\beta - \beta_c| / \beta_c$
on the largest lattice ($N=2500$) as a consistency check.
If we fit
\begin{equation}
C \simeq A' + B' t^{- \alpha}
\end{equation}
we again get a good value $\alpha = -1.1(1)$ because of the extra
constant $A'$. The fit to $\beta$
\begin{equation}
M \simeq M_0' t^{\beta}
\end{equation}
is even more mediocre than the finite size scaling 
fit for $\beta / \nu D$,
giving a wide range of values with similar goodness of fit
depending on the points deleted from the fit.
The direct fit to $\gamma$
\begin{equation}
\chi \simeq \chi_0' t^{- \gamma},
\end{equation}
on the other hand, is very good: $\gamma = 2.07(2)$. 

For completeness we list our fitted exponents
and critical temperature for the ferromagnetic simulations
below along with the analytical values for
the critical exponents
\begin{center}
\begin{tabular}{|c|c|c|c|c|c|c|c|c|c|} \hline
 & $\alpha$  & $\beta$  & $\gamma$ & $\nu D$ &
$\alpha/\nu d$ & $\beta/\nu d$ & $\gamma/\nu d$ & $\beta_c$   \\[.05in]
\hline
$measured$& -1.1(1)  & $\ldots$ & 2.07(2) & 3.2(3) &
-0.35(2)  & 0.075(2)     & 0.72(2)  & 0.398(2)       \\[.05in]
\hline
$theory$& -1  & 1 / 2  & 2  & 3  &
     -1/3   & 1 / 6      &   2 / 3   & $\ldots$     \\[.05in]
\hline
\end{tabular}
\end{center}
\vspace{.1in}
\centerline{Table 1: Measured 
and Theoretical values of $\alpha, \beta, \gamma, \nu D$.}
\centerline{(We have not included the very poor direct fit to $\beta$.)}
\vspace{.1in}

We close with a few observations on the efficiency of the 
Monte-Carlo algorithms. Previous work on triangulations and
$\phi^3$ graphs has shown that although cluster algorithms
greatly reduce critical slowing down for energy observables
the magnetization still has quite large values of the
dynamical critical exponent $z$ \cite{3a}, defined
through the scaling of the appropriate autocorrelation
time $\tau$
\begin{equation}
\tau \simeq N^{z / D}.
\end{equation}
For the ferromagnetic simulations 
with the Wolff cluster algorithm we find $z_M / D = 0.39(3)$
for the magnetization,
and $z_E / D = 0.23(4)$ for the energy.
These values are unlikely to be particularly accurate
because of the small lattice sizes, but the value for the
energy is unusually large compared with previous $\phi^3$
and triangulation simulations. A glance at the metropolis
simulations for the antiferromagnet shows that the 
cluster algorithm autocorrelation
times are an order of magnitude better for the magnetization
and a factor of two or three better for the energy on the 
$N=2500$ graphs, so they still offer an improvement over
a local algorithm in spite of the 
anomalously large $z_E$. 

It was suggested in \cite{3a} that the bottlenecks that appear
in the baby universe structure of the $\phi^3$ graphs 
and triangulations impeded cluster
formation and was responsible for this relatively poor performance.
Although we have not investigated the baby-universe distribution
of our quadrangulations there is no reason to suppose it is
significantly different from triangulations as we are clearly
simulating the same continuum theory if the Ising results
are to be believed. We would therefore expect the same 
bottlenecks to be present on our quadrangulations, causing the same
obstacles to cluster growth. By the same token, it would be equally
easy to perform baby universe surgery \cite{2c} on a quadrangulation
to help alleviate the problem, the only difference from a triangulation
being that the minimum neck is now of length four rather than three.

Other geometrical features such as the flip acceptance
as a function of temperature and the cluster distributions look
qualitatively similar to earlier $\phi^3$ and triangulation
simulations.
As an additional measure to improve our measurements of
the magnetization critical exponent
it would be a relatively easy matter to include degenerate
quadrangulations in the class of graphs we simulate
in order to improve the convergence to the continuum limit.

\section{Conclusions}
Modulo the usual
rather poor fits for the magnetization
exponent $\beta$ the measured critical exponents
for ``square gravity''
are consistent with the KPZ/DDK values, thus providing further
explicit
numerical support for the notion of universality on dynamical lattices
as far as the ferromagnetic transition is concerned.
As a by-product of the simulations we have 
also found the critical temperature for the non-degenerate
quadrangulations that we considered here $\beta_c = 0.398(2)$.
This translates to the dual value $\beta_c  \simeq  0.486$
on non-degenerate $\phi^4$ graphs, to be compared with
$\beta_c \simeq  0.693$ for $\phi^4$ graphs in which the various
bubbles etc. are allowed. 

It is a relatively simple matter to simulate
the dual $\phi^4$ model as well, but here of course we would
no longer find an antiferromagnetic transition.
We need to consider loosely
packed lattices on which Neel order is possible
to observe the antiferromagnetic transition
on fixed lattices, which remains the case for dynamical lattices.
On dynamical lattices we are summing over some class of 
loosely packed lattices, which we 
implement in the simulations with 
a compatible flip move.
From the matrix model and the simulations
it is clear that the antiferromagnetic transition
survives its coupling to 2d gravity
on dynamical quadrangulations
and retains the same exponents as the ferromagnetic transition.
The antiferromagnet thus provides an amusing example of 
Neel order surviving on what is essentially a fluid
random lattice.

Extending the approach here to investigate other polygonizations
is also not difficult. Exploding higher order vertices
into $\phi^3$ vertices in such models allows a derivation
of the appropriate flip moves and weighting factors.

\section{Acknowledgements}
CFB and DAJ were supported in part by NATO grant CRG910091.
CFB is supported by 
by DOE under
contract DE-FG02-91ER40672, by NSF Grand Challenge Applications
Group Grant ASC-9217394 and by NASA HPCC Group Grant NAG5-2218.
DAJ is also supported in part by EC grant CHRXCT930343.

\bigskip
\centerline{\bf Figure Captions}
\begin{description}
\item{Fig.1}
The two possible flip moves on adjacent squares in DQRS both preserve the ``checkerboard'' order
of the ground state. Up spins are shown as solid dots and down spins as open dots. 
\item{Figs.2} 
The possible ways of ``exploding'' the $\phi^4$ diagram dual to two adjacent squares. The $u$
propagators are represented as dotted lines and the $\phi$ propagators as solid lines.
\item{Figs.3} The resulting $\phi^4$ flip moves once the diagram has been pieced
together again after carrying out all the possible  flips in the exploded model. Both orientations
of the central link occur with equal probability in both diagrams. Going back to the direct lattice
gives the move depicted in Fig.1
\item{Fig.4}
The energy for both the ferromagnetic and antiferromagnetic models on
the largest lattices simulated ($N=2500$).
\item{Fig.5}
The magnetization for both the ferromagnetic and antiferromagnetic models
on the largest lattices.
\item{Fig.6}
The specific heat $C$ for the ferromagnet.
\item{Fig.7}
The susceptibility $\chi$ for the ferromagnet.
\end{description}
\vfill
\eject
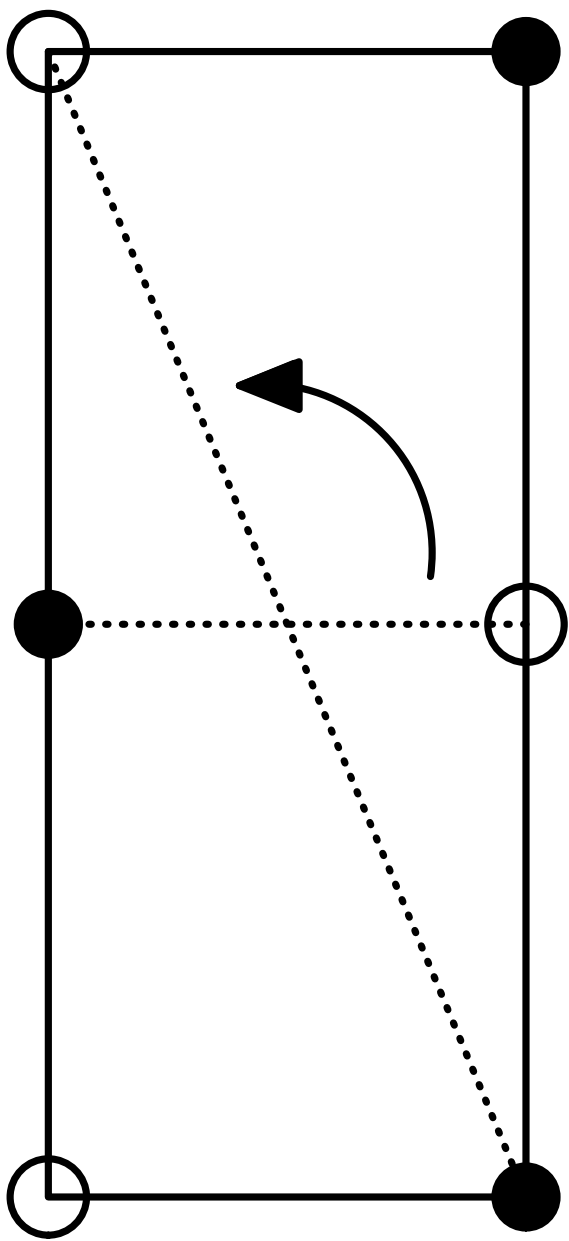
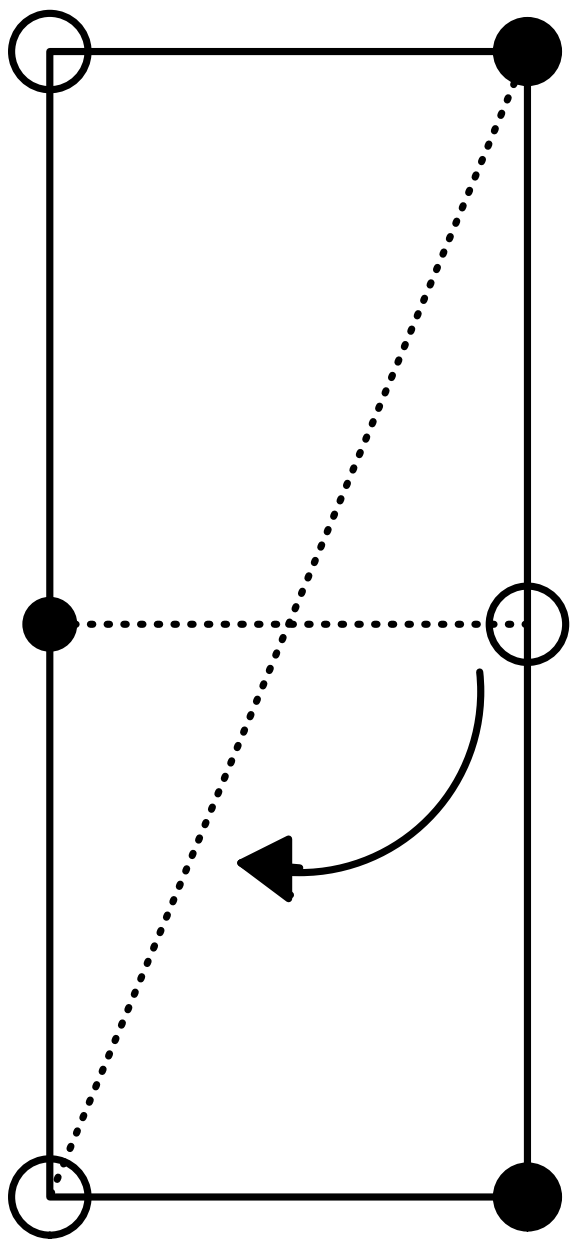
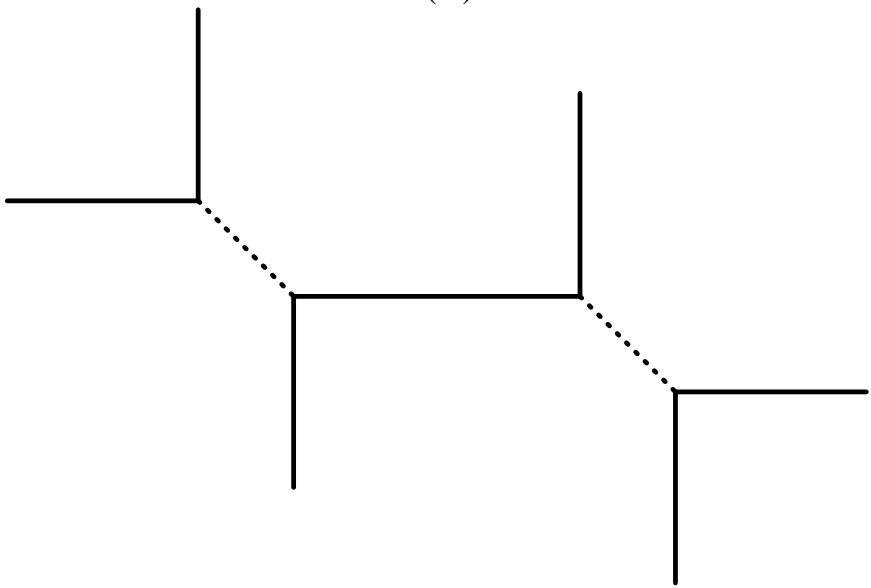
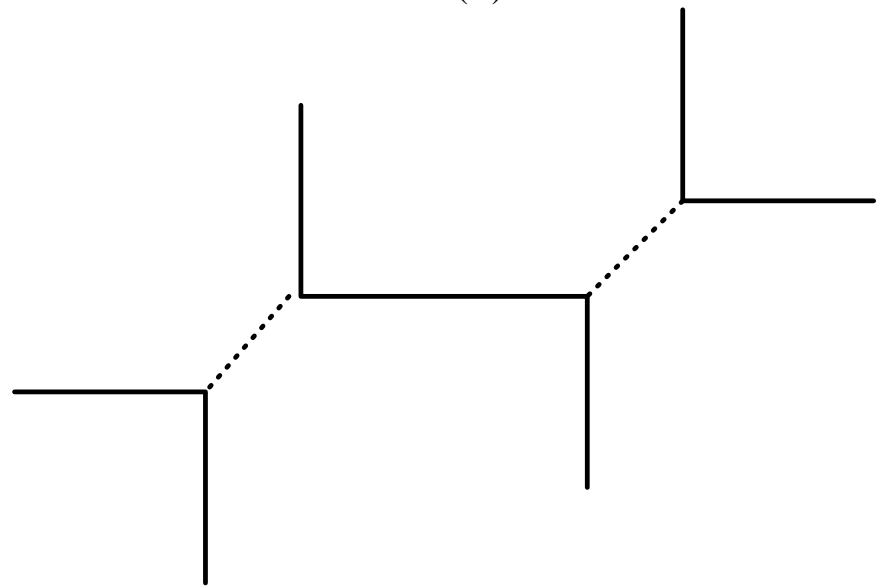
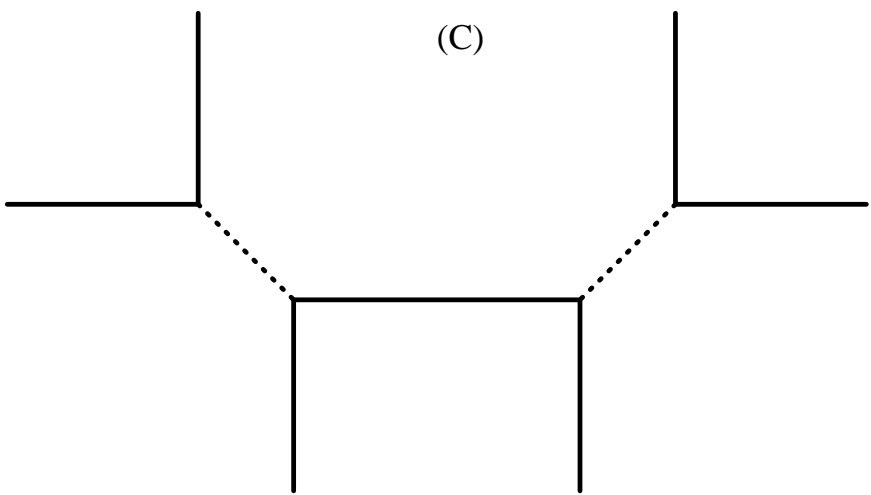
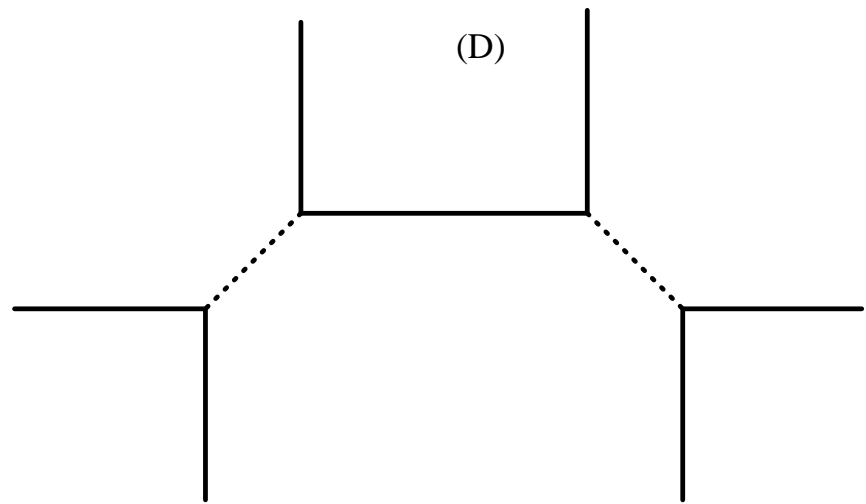
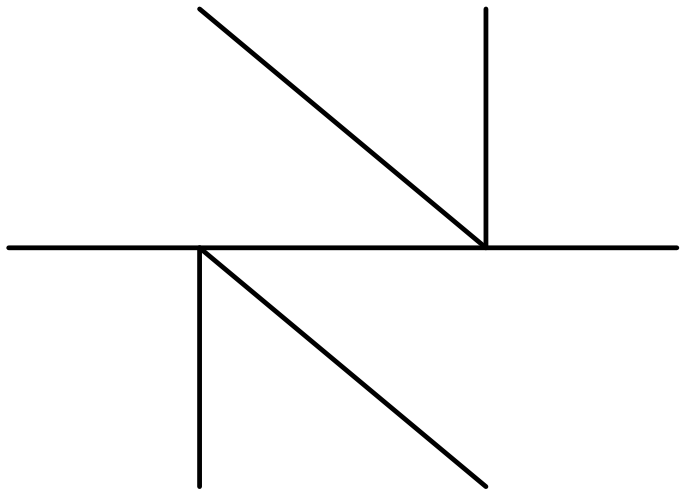
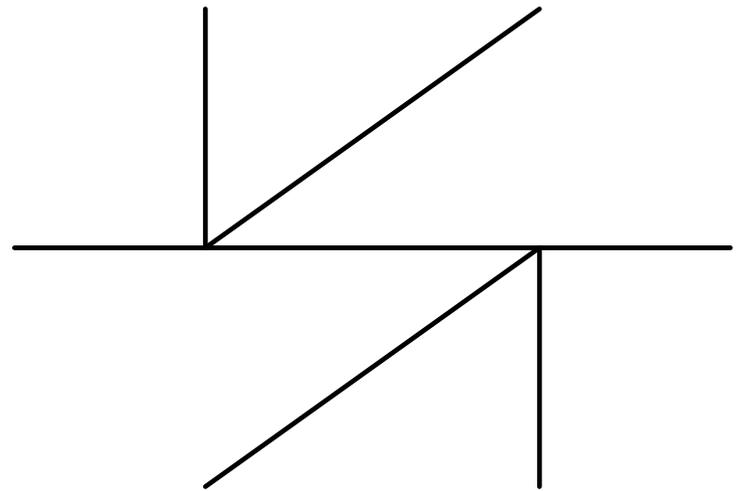
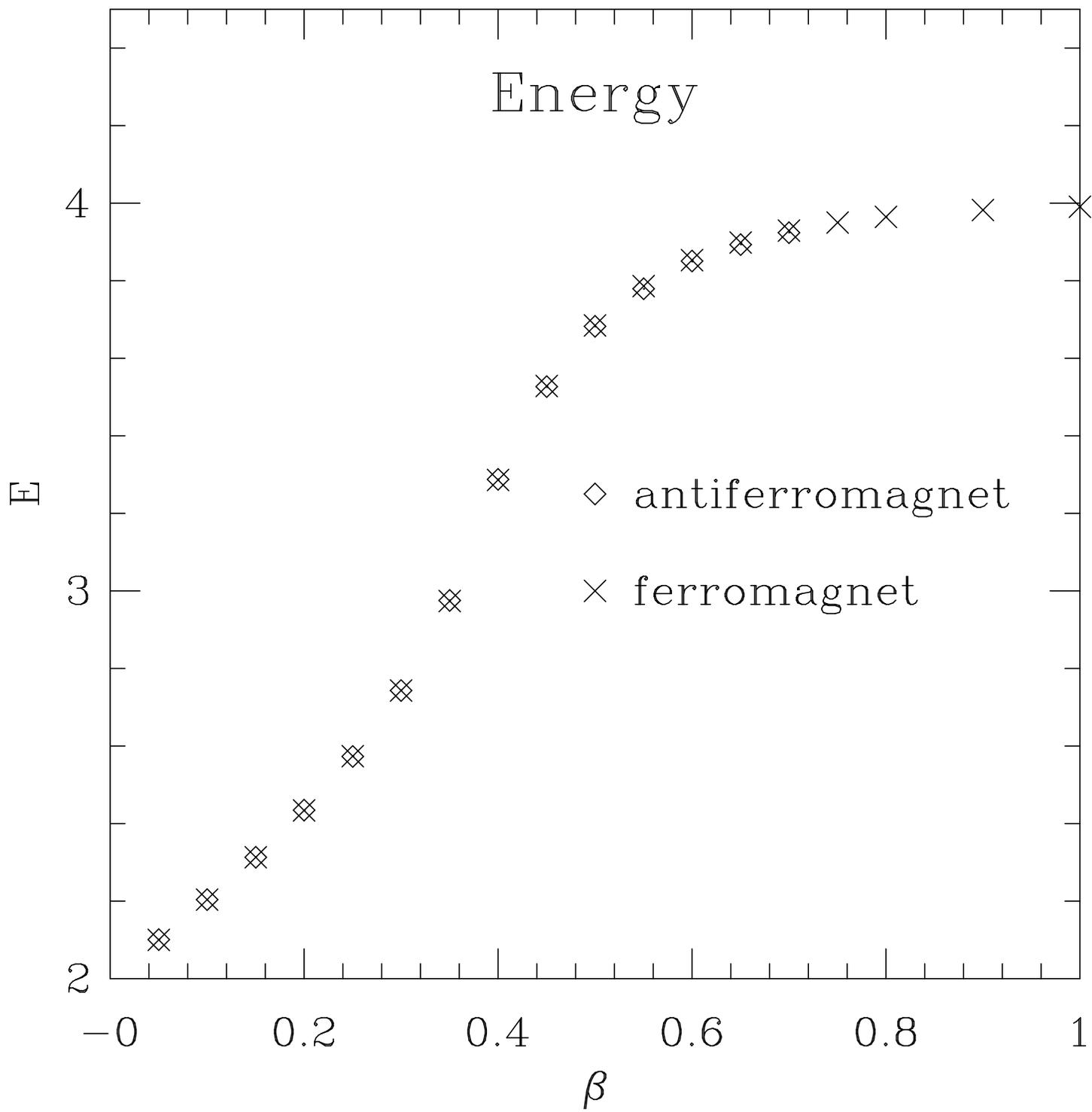
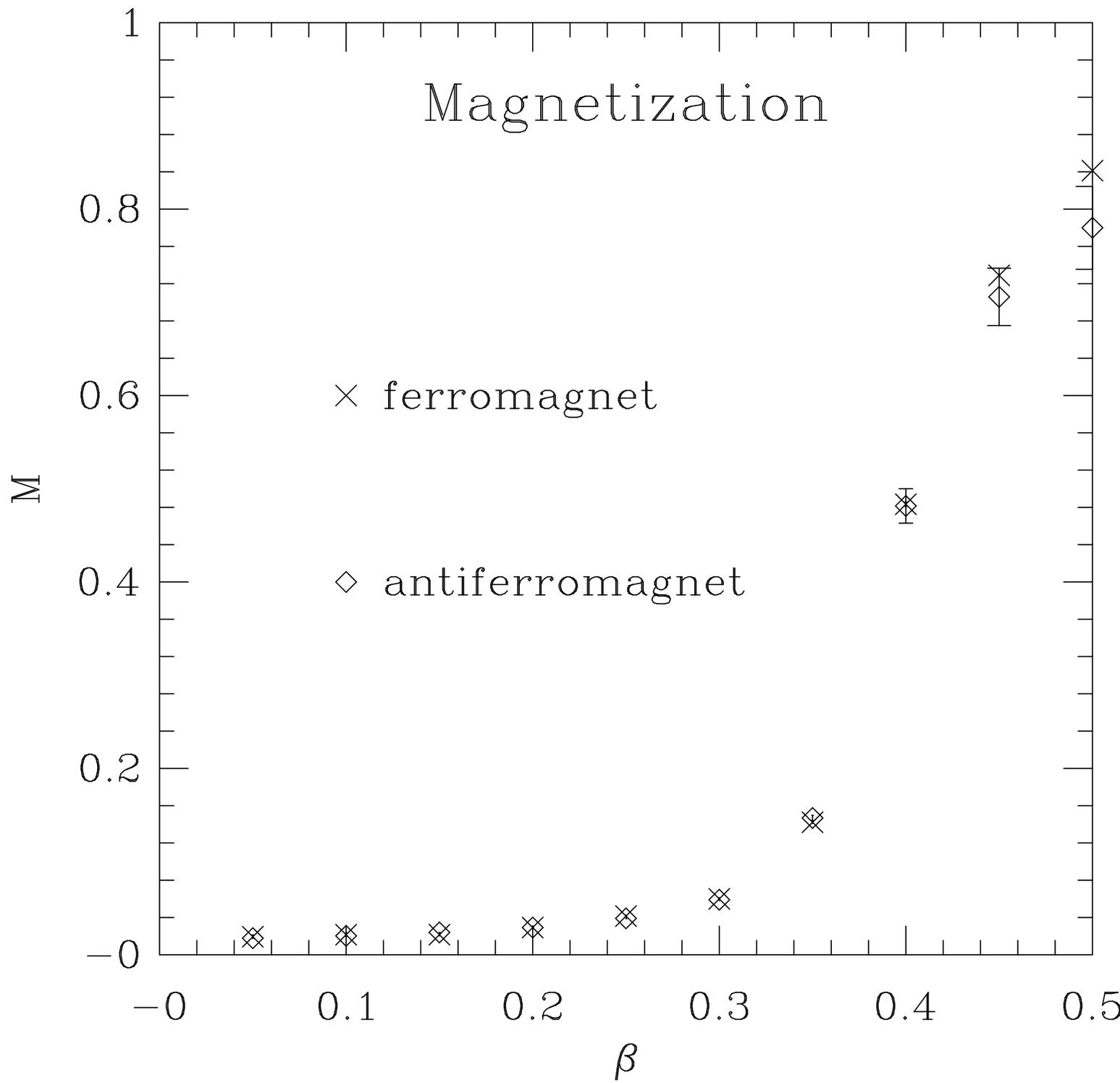
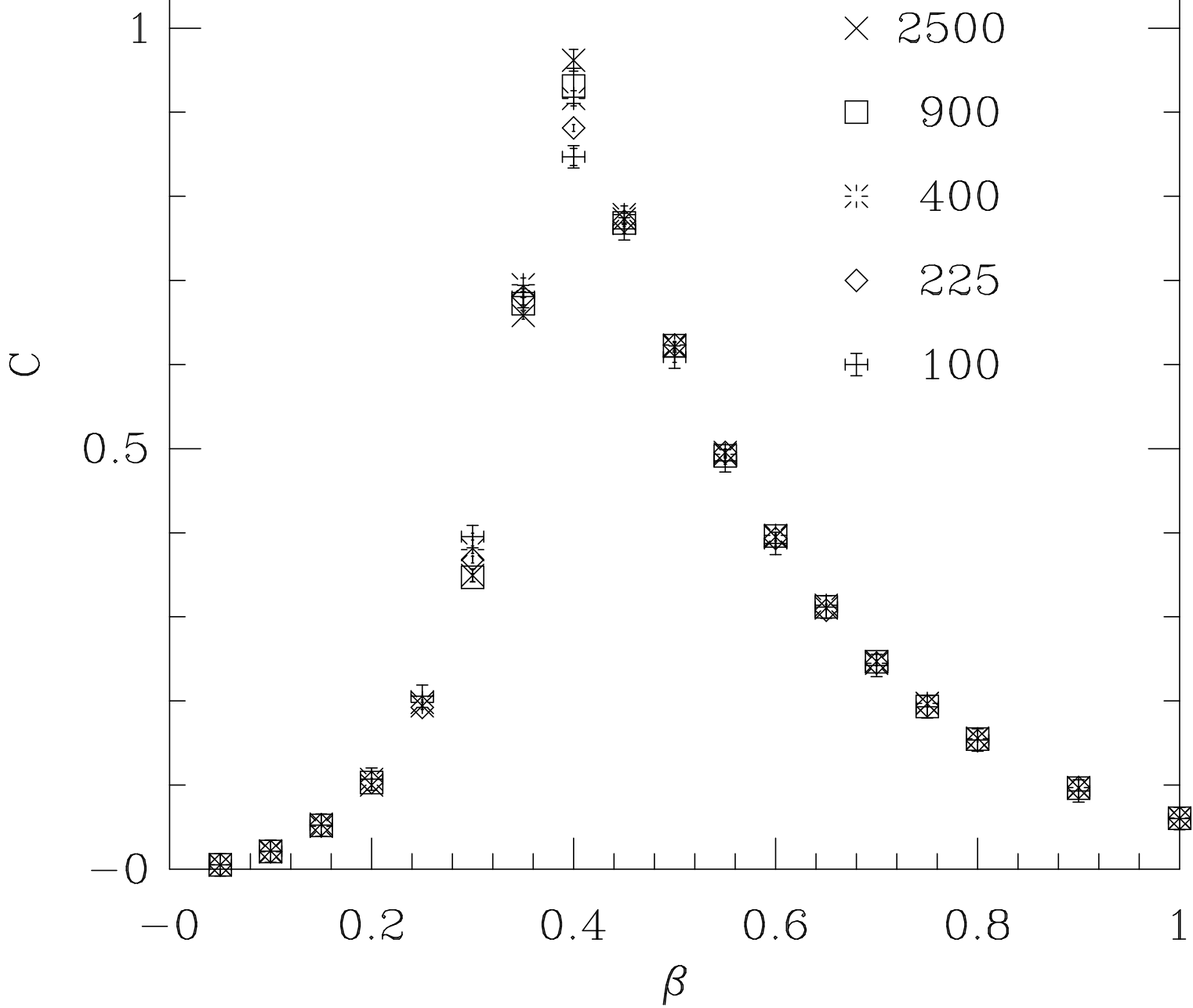
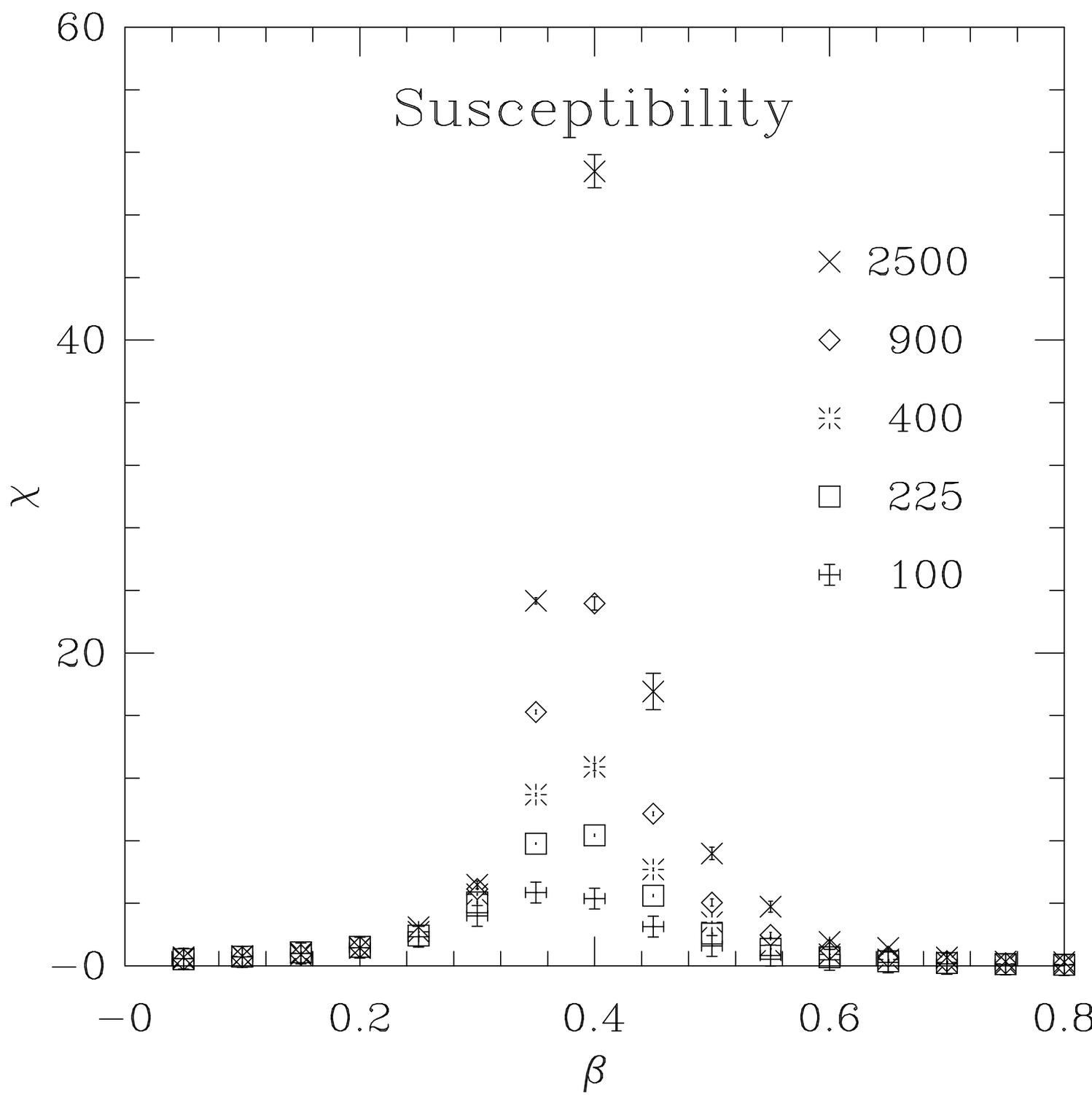
\end{document}